\documentclass{elsarticle}
\usepackage{amsmath}
\usepackage{latexsym}

\def\dfrac{\displaystyle\frac}
\def\llc{\left\lceil}
\def\rrc{\right\rceil}
\def\Ref#1{(\ref{#1})}
\def\eps{\varepsilon}

\newtheorem{theorem}{Theorem}
\newtheorem{remark}{Remark}

\newproof{proof}{Proof}

\renewcommand{\textfraction}{0.01}

\begin{document}
\begin{frontmatter}

\title{Allocation of seats in the European Parliament and a~degressive proportionality.}
\author{Jan~Florek}
\address{Institute of Mathematics,\\
University of Economics,\\
ul. Komandorska 118/120\\
53--345 Wroc{\l}aw, Poland}

\begin{abstract}
Distribution of seats in The European Parliament postulated by Treaty of Lisbon should be degressively  proportional. The meaning of degressively  proportional concept can be found in two principles annexed to the draft of European Parliament resolution. The first, referred as the principle of fair division, states that \textit{,,the larger the population of a Member State, the greater is entitlement to a large number of seats''}. The other condition, referred to as the principle of relative proportionality, holds that \textit{,,the larger the population of a country, the more inhabitants are represented by each of its Members of the EU''}. We postulate a clear and fair method which determines uniquely a distribution of seats in the European Parliament which fulfil the requirements of degressive proportionality.
More generally, let $l_i$ be any non-increasing sequence of real positive numbers.
We say that a sequence of natural numbers $m_{i}$ is degressively proportional with respect to the sequence $l_{i}$, if $m_{i}$ and $l_{i}/m_{i}$ are
non-increasing sequences. Our method can be instrumental in uniquely determining a degressively proportional sequence $m_{i}$
with respect to $l_{i}$ which fulfils given conditions.
\end{abstract}


\end{frontmatter}

\section{Introduction}
The European Parliament is one of the most important institutions of the European Union based on representations of members states. Principles of seats distribution in The EP have changed with subsequent EU enlargement stages. Due to large distribution of population between individual member states, no proportional method can be employed in seat distribution.
Therefore another approach to apportionment was postulated. The postulate was expressed in Article 9a paragraph 2 of the Treaty of Lisbon. The article states that:

\textit{,,The European Parliament shall be composed of representatives of the Union's citizens. They shall not exceed seven hundred and fifty in number, plus the President. Representations of citizens shall be degressively proportional, with a minimum threshold of six members per Member State. No Member State shall be allocated more than ninety-six seats''.}(Treaty of Lisbon \cite{lisbon}).

\looseness-1
The meaning of the concept of degressive proportionality can be found in two principles annexed to the draft of European Parliament resolution. The first, referred as the principle of fair division, states that \textit{,,the larger the population of a Member State, the greater is entitlement to a large number of seats''}. The other condition, referred to as the principle of relative proportionality, holds that \textit{,,the larger the population of a country, the more inhabitants are represented by each of its Members of the EU''.} (Lamassoure and Severin \cite{lamssoure}). A formal approach to the definition of degeressive proportionality was studied by Ramirez-Palmarez-Marquez \cite{ramirez} and  {\L}yko-Cegie{\l}ka-Dniestrza\'{n}ski-Misztal \cite{lyko}. Let $n$ represent the number of Members States, $l_i$ - the population of the $i$-th member, and $m_i$ - the number of mandates offered to the Member State. Suppose that $l_1 > l_2 > \ldots > l_n$. Then the sequence $m_i$ is degressively proportional with respect to the sequence $l_i$ if it is non-increasing and satisfies the following condition:
\[
\frac{l_{1}}{m_{1}} > \frac{l_{2}}{m_{2}} > \ldots > \frac{l_{n}}{m_{n}} \, .
\]
The present composition of the European Parliament does not satisfy the principles of the degressive proportionality. Distribution of seats in the European Parliament postulated by Committee on Constitutional Affairs members Lamassoure and Severin does indeed fulfill the requirements of degressive proportionality. The main problem is to find a clear and fair method (acceptable for all Member States) which determines uniquely a sequence $m_i$ degressively  proportional with respect to $l_i$.

For a real number $x$ we denote by $\llc{x}\rrc$ the least integer $\geq x$. All the following sequences are sequences of length $n$ or $n-1$,
where $n\ge2$ is a fixed number.
The definition of degressive proportionality can be slightly extended as follows.
Let $l_i$ be any fixed non-increasing sequence of real positive numbers.
We say that a sequence of natural numbers $m_{i}$ is
\textit{degressively proportional with respect to the sequence} $l_{i}$,
if $m_{i}$ is non-increasing and
\begin{equation}
\label{eq1}
\frac{m_{1}}{l_{1}} \leq  \frac{m_{2}}{l_{2}} \leq \ldots \leq \frac{m_{n}}{l_{n}}.
\end{equation}
It Theorem~1(a) we prove that a sequence of natural numbers $m_{i}$ is
degressively proportional with respect to $l_{i}$
if and only if it is defined inductively
\begin{equation}
\label{eq2}
\left\{
\begin{array}{l}
m_{1}= M, \\
m_{i}= \min \left(\llc \dfrac{m_{i-1}l_{i}}{l_{i-1}}+a_{i}\rrc,m_{i-1}\right),\mbox{ for $2 \leq  i \leq n$},
\end{array}
\right.
\end{equation}
for some sequence $a_i \geq 0$ and a constant $M \in\mathbf{N}$. Note that a sequence $M_i$ defined inductively in Theorem~1(b) is degressively proportional with respect to $l_{i}$ and is smaller, then any other such sequence with the first element $\geq M$. The pair $(M,\{a_i\})$ will be called \textit{the initial condition for the sequence} $m_{i}$.

Fix constants $M, Y \in\mathbf{N}$ such that $Y \geq \sum{M_i}$.
We suggest the following method of the choice of a sequence $m_i$ degressively  proportional with respect to $l_i$ and satisfying the inequalities
\begin{equation}
\label{eq3}
\sum{m_i} \leq Y \quad \mbox{and} \quad m_1 = M \, .
\end{equation}
Fix a sequence $a_i \geq 0$. For every $c\ge0$, let $m_i(c)$ denote the degressively proportional sequence with respect to $l_i$, with the initial condition $(M,\{ca_i\})$. In Theorem~2(a) we prove that all the functions $m_{i}\colon{c\mapsto{m_i(c)}}$, $1 \le  i \leq  n$, and the function $\Phi = \sum m_i$, are left-continuous and non-decreasing on the positive half line $[0,\infty)$.
Hence, there exists a positive number $c_Y$ such that
$$
\Phi(c_Y) = \max \{ \Phi(c) : \Phi(c) \leq Y \}.
$$
We postulate the choice of the sequence $m_i(c_Y)$ dominating all of other sequences degressively proportional with respect to
$l_{i}$, with initial condition $(M,\{ca_i\})$, for every $c > 0$,
and satisfying inequalities \Ref{eq3}.

In Theorem~3(b) we determine the constant $\delta$ such that
$\Phi(\delta)$ is the largest value of the function $\Phi$. If $Y \in [\Phi(0),\Phi(\delta)]$, then Theorem~4(b) provides the lower bound of the value $\Phi(c_Y)$. This implies
$$
Y-(n-2) \leq \sum{m_i}(c_Y) \le  Y.
$$
The problem of computing the number $c_{Y}$ is equivalent
to finding some points (not necessarily the discontinuity points)
at which the function $\Phi$ takes its successive values.
We present two methods which lead to finding such points:
\begin{itemize}
\item[(a)]
If $\Phi(c_0)$ is not the largest value of $\Phi$,
then $c_0+~\beta(c_0)$ is the first discontinuity point
belonging to $[c_0,\infty)$, where $\beta(c_0)$
is the constant described in Theorem~5.
\item[(b)]
For every $c \geq 0$, the function $\Phi$
has in $[c,c+\gamma(c)]$ at most one point of discontinuity, where $\gamma(c)$
is the constant of Theorem~6.
If $\Phi(c) < \Phi(c+\gamma(c))$, then $\Phi(c)$ and $\Phi(c +\gamma(c))$ are consecutive values of $\Phi$.
We are able, therefore, to find consecutive values of $\Phi$
without knowing of the discontinuity points of $\Phi$.
\end{itemize}
Hence, if $\Phi(c)$ is not the largest value of the function $\Phi$, then
$$\Phi(c) \mbox{ and } \Phi(c+\beta(c)+\gamma(c+\beta(c)))$$
are consecutive values of $\Phi$ (see Remark 4). For a real number $x$ we denote by ${\lfloor x\rfloor}_k$ the decimal representation of $x$ up to $k$ digits after the decimal point. It is convenient to choose $k$ such that
$$
{\lfloor {(1-l_{i}/l_{i-1})/a_{i}} \rfloor}_k > 0, \mbox{ for } {a_{i}}\neq 0 \mbox{ and } l_{i} \neq l_{i-1},
$$
(cf.~the constants $\gamma_3$ and $ \gamma$ of Theorem~6). Using  Excel we can easily compute
 $$\Phi(c) \mbox{ and } \Phi(c+{\lfloor\beta(c) \rfloor}_k+{\lfloor \gamma(c+{\lfloor\beta(c) \rfloor}_k)\rfloor}_k)$$
which are practically different (accordingly, consecutive values of $\Phi$).

\looseness-1
Let us come back to the initial problem of distribution of
sets in the European Parliament. In this case $n=27$,
$l_1 > l_2 > \ldots > l_{27}$ is a sequence of populations of Member States, and $M = 96$ is the number of mandates offered to Germany. The number $a_i$, $2 \leq i \leq n$, can be regarded as the degree of preference of the $i$-th Member State. Since a natural intention of the European community is to offer a fair representation to all members, we consider the case when $a_i$ is a constant sequence, say $a_i \equiv 1$, as reflecting this intention.
For each $c \geq 0$, let $m_i(c)$ be a degressively proportional
sequence with respect to $l_i$,  with the initial condition $(96,\{c\})$.

In the columns of Table 1 we present the values
of the functions $m_{i}\colon{c\mapsto{m_i(c)}}$, $1 \le  i \leq  27$,
and the function $\Phi = \sum m_i$ at the points
$$
c_1 = 1.11, c_2 = 1.140625, c_3 = 1.25731913,
$$
$$
c_4 = 1.5, c_5 = 1.555, c_6 = 1.6, c_7 = 1.7 \,,
$$
respectively.
Notice that
$$
(\sum m_i(c_k), m_{27}(c_k)) = (736,5), (751, 5),(757, 6), \quad \mbox {for k = 1, 2, 3 }.
$$
\def\DE{Germany}
\def\FR{France}
\def\GB{Gr. Britain}
\def\IT{Italy}
\def\ES{Spain}
\def\PL{Poland}
\def\RO{Romania}
\def\NL{Netherlands}
\def\GR{Greece}
\def\PT{Portugal}
\def\BE{Belgium}
\def\CZ{Czech Rep.}
\def\HU{Hungary}
\def\SE{Sweden}
\def\AT{Austria}
\def\BG{Bulgaria}
\def\DK{Denmark}
\def\SK{Slovak Rep.}
\def\FI{Finland}
\def\IE{Ireland}
\def\LT{Lithuania}
\def\LV{Latvia}
\def\SI{Slovenia}
\def\EE{Estonia}
\def\CY{Cyprus}
\def\LU{Luxemburg}
\def\MT{Malta}

\def\1#1,#2,#3,#4,#5,#6,#7,#8;{#1&#2&#3&#4&#5&#6&#7&#8\\
\hline}
\begin{table}
\begin{tabular}{l*7{c}}
\hline
\1Country,$m_i(c_1)$,$m_i(c_2)$,$m_i(c_3)$,$m_i(c_4)$,$m_i(c_5)$,$m_i(c_6)$,$m_i(c_7)$;
\1\DE,96,96,96,96,96,96,96;
\1\FR,75,75,75,75,75,75,76;
\1\GB,74,74,74,74,74,74,75;
\1\IT,74,74,74,74,74,74,75;
\1\ES,57,57,57,57,57,57,58;
\1\PL,51,51,51,52,52,52,53;
\1\RO,30,31,31,31,32,32,32;
\1\NL,24,25,25,25,26,26,26;
\1\GR,18,19,19,19,20,20,20;
\1\PT,18,19,19,19,20,20,20;
\1\BE,18,19,19,19,20,20,20;
\1\CZ,18,19,19,19,20,20,20;
\1\HU,18,19,19,19,20,20,20;
\1\SE,18,19,19,19,20,20,20;
\1\AT,18,19,19,19,20,20,20;
\1\BG,18,19,19,19,20,20,20;
\1\DK,14,15,15,15,16,16,16;
\1\SK,14,15,15,15,16,16,16;
\1\FI,14,15,15,15,16,16,16;
\1\IE,13,14,14,14,15,15,15;
\1\LT,12,13,13,13,14,14,14;
\1\LV,10,10,11,11,11,12,12;
\1\SI,10,10,11,11,11,12,12;
\1\EE,8,8,9,9,9,10,10;
\1\CY,6,6,7,7,7,8,8;
\1\LU,5,5,6,6,6,7,7;
\1\MT,5,5,6,6,6,7,7;
\1Total,736,751,757,758,773,779,784;
\end{tabular}
\caption{}
\end{table}

Using  Excel and methods (a) and (b), we can prove that the numbers
$$ 736, 751, 757, 758, 773, 779, 784$$
are all values of the function $\Phi = \sum m_i$
which are in  the interval $[736, 784]$ (see Examples).

\section{Main result}
Let $l_i$ be a non-increasing sequence of real positive numbers.
In Theorem~1 we characterize all degressively proportional sequences with
respect to $l_{i}$.

\begin{theorem}
Let  $l_{i}> 0$ be any non-increasing sequence.
\begin{itemize}
\item[(a)]
A sequence of natural numbers $m_i$ is degressively proportional with respect to   $l_{i}$ if and only if it is defined inductively by \Ref{eq2} for some sequence $a_i \geq 0$ and a constant $M \in\mathbf{N}$.
\item[(b)]
The following sequence
\[
\left\{
\begin{array}{l}
M_{1}= M, \\
M_{i}= \llc\dfrac{M_{i-1}l_{i}}{l_{i-1}}\rrc, \mbox{ for $2 \le  i \leq n$}.
\end{array}
\right.
\]
is a non-increasing \textit{minorant}, that is, it is smaller than any other sequence with the first element $\geq M$ which satisfies condition \Ref{eq1}.
\end{itemize}
\end{theorem}

\begin{proof}
If $m_{i}$  is a sequence  of natural numbers defined inductively by \Ref{eq2}, then $m_{i} \geq m_{i-1}l_{i}/l_{i-1}$. Hence, $m_i$ satisfies \Ref{eq1}. If $m_{i}$ is a non-increasing sequence  of natural numbers which satisfies \Ref{eq1}, then $a_{i}= m_{i}-m_{i-1}l_{i}/l_{i-1}\geq 0$. Hence, $m_{i}$ is defined inductively by \Ref{eq2}.

From $M_{i-1}l_{i}/l_{i-1}\leq M_{i-1}$, we see that $M_{i}$ is non-increasing.  Let $v_{i}$ be a sequence  of natural numbers which satisfies \Ref{eq1}, and $v_{1}\ge  M$. We prove (b) by induction. Assume that $M_{i}\leq v_{i}$. Then,
$$
\frac{M_{i}}{l_{i}} \leq \frac{v_{i}}{l_{i}} \leq \frac{v_{i+1}}{l_{i+1}}.
$$
Hence, $M_{i+1} = \llc M_{i}l_{i+1}/l_{i}\rrc \leq v_{i+1}$, which concludes (b).
\qed
\end{proof}

\begin{remark}
Without loss of generality we can assume (see \Ref{eq2})
that the sequence $a_{i}$ satisfies the following implication:
\begin{equation}
\label{eq4}
\mbox{For every $2 \le  i \leq n$, if $l_{i}= l_{i-1}$  then  $a_{i}= 0$}.
\end{equation}
\end{remark}

\begin{theorem}
Let  $l_{i}> 0$ be any non-increasing sequence, $M \in\mathbf{N}$, and $a_{i}\ge  0$. For every $c\ge0$, let $m_i(c)$ denote a degressively proportional
sequence with respect to $l_i$, with the initial condition $(M,\{ca_i\})$. Set \begin{equation}
\label{eq5}
A_{i}(c) = \frac{m_{i-1}(c)l_{i}}{l_{i-1}}+ ca_{i} \, ,
\end{equation}
for $2 \le  i \leq n$, and $c \geq 0$.

Then all the functions  $m_{i}\colon{c\mapsto{m_i(c)}}$, and the function $\Phi  =\sum m_i$, are left-continuous and non-decreasing on the positive half line $[0,\infty)$.
\end{theorem}

\begin{proof}
It is easily seen that
 all the functions $m_{i}$, $2 \le  i \leq  n$, and the function $\Phi$  are non-decreasing. We only need to show the following implications:

(i) for every  $1 \le  i < n$, if $m_{i}(c-\eps ) = m_{i}(c)$ for sufficiently small $\eps  > 0$, then $\llc A_{i+1}(c-\eps)\rrc = \llc A_{i+1}(c)\rrc$ for another sufficiently small $\eps  > 0$.

(ii) for every  $1 \le  i < n$ and $0 \leq c \leq d$, if $\llc A_{i+1}(c)\rrc = \llc A_{i+1}(d)\rrc$ and $m_{i}(c) = m_{i}(d)$, then $m_{i+1}(c) = m_{i+1}(d)$.

Proof (i). We proceed by induction on $i$.
Assume that $m_{i}(c-\eps ) = m_{i}(c)$ for sufficiently small $\eps > 0$.
Then for another sufficiently small $\eps  > 0$
we obtain
$$
\llc A_{i+1}(c)\rrc -1 < A_{i+1}(c-\eps ) = A_{i+1}(c) - \eps  a_{i+1}
\le \llc A_{i+1}(c)\rrc.
$$
Hence, $\llc A_{i+1}(c-\eps)\rrc = \llc A_{i+1}(c)\rrc$.

Proof (ii). We proceed by induction on $i$. Assume that $\llc A_{i+1}(c)\rrc = \llc A_{i+1}(d)\rrc$ and $m_{i}(c) = m_{i}(d)$. Then,
$$
m_{i+1}(c) = \min \left(\llc A_{i+1}(c)\rrc, m_{i}(c)\right)
 = \min \left(\llc A_{i+1}(d)\rrc, m_{i}(d)\right) = m_{i+1}(d),
$$
which completes the proof.
\qed
\end{proof}

\begin{theorem}
Let  $l_{i}> 0$ be any non-increasing sequence, $M \in\mathbf{N}$, and $a_{i}\ge  0$.
For every $c\ge0$, let $m_i(c)$ denote a degressively proportional
sequence with respect to $l_i$, with the initial condition $(M,\{ca_i\})$.
\begin{itemize}
\item[(a)]
Suppose that $c_0 \ge  0$ satisfies the following condition:
$$
\mbox{For every  $2 \le  i \leq n$ , \,  $m_{i}(c_0) = m_{i-1}(c_0)$ or $a_{i} = 0$ .}
$$
Then all functions $m_{i}\colon{c\mapsto{m_i(c)}}$, $1 \le  i \le  n$, and the function  $\Phi  =\sum m_i$ are constant on the half line $[c_0,\infty)$.
\item[(b)]
Set
$$
\delta = {M}\max\left\{ \frac1{a_{i}}\left(1-\frac{l_{i}}{l_{i-1}}\right):{a_{i}}\neq 0 \right\}.
$$
The function  $\Phi$ takes all its values in the interval $[0,\delta]$.
\end{itemize}
\end{theorem}

\begin{proof}
Fix $c\geq c_0$. Let $A_{i}(c)$,  $2 \le  i \leq n$, be the sequence defined by  \Ref{eq5}. We proceed by induction on $i$. Assuming $m_i(c) = m_i(c_0)$, we will prove it for $i+1$.  If $m_{i+1}(c_0) = m_{i}(c_0)$, then
$$
m_{i+1}(c)=\min\left(\llc A_{i+1}(c)\rrc,m_{i}(c)\right)
=\min\left(\llc A_{i+1}(c)\rrc,m_{i+1}(c_0)\right)=m_{i+1}(c_0).
$$
The last equality follows from $\llc A_{i+1}(c)\rrc\geq\llc A_{i+1}(c_0)\rrc\geq m_{i+1}(c_0)$.
If $a_{i+1}=0$, then $\llc A_{i+1}(c)\rrc=\llc A_{i+1}(c_0)\rrc$. Hence, $m_{i+1}(c) = m_{i+1}(c_0)$.

We turn to condition (b). If $a_{i} \neq 0$, then, by the definition of the number $\delta$, we have
$$
\delta \geq  \frac{M}{a_{i}}\left(1-\frac{l_{i}}{l_{i-1}}\right) \geq  \frac{m_{i-1}(\delta)}{a_{i}}\left(1-\frac{l_{i}}{l_{i-1}}\right) \, .
 $$
Hence,
$$
A_{i}(\delta) = \frac{m_{i-1}(\delta)l_{i}}{l_{i-1}}+ \delta a_{i} \geq m_{i-1}(\delta) \, .
$$
Accordingly,
$$
m_{i}(\delta) = \min \left(\llc A_{i}(\delta)\rrc, m_{i-1}(\delta)\right) = m_{i-1}(\delta).
$$
Hence, by (a), the function $\Phi$ is constant on the half line $[\delta,\infty )$.
\qed
\end{proof}

\begin{theorem}
Let  $l_{i}> 0$ be any non-increasing sequence,
 $M \in\mathbf{N}$, and suppose that  $a_{i}\ge  0$ is different from the zero sequence $(0,\ldots,0)$ satisfying the condition \Ref{eq4}. For every $c\ge0$, let $m_i(c)$ denote a degressively proportional sequence with respect to $l_i$, with the initial condition $(M,\{ca_i\})$.
\begin{itemize}
\item[(a)]
Then,
$$
\alpha = \min\left\{\frac1{a_{i}}\left(1-\frac{l_{i}}{l_{i-1}}\right): {a_{i}}\neq 0  \right \}>0 \, .
$$
For every $2 \le  i \leq n$, and $c \geq 0$,
$$
 m_{i}(c+\alpha ) \le  m_{i}(c) + 1.
$$
\item[(b)]
For every natural number $Y \in[\Phi(0), \Phi(\delta)]$, there is a point $x \in  [0, \delta]$  such that
$$
Y -(n-2) \le  \Phi (x) \le  Y.
$$
\end{itemize}
\end{theorem}

\begin{proof}
Since $a_{i}$  is different from the zero sequence and satisfies implication \Ref{eq4}, then $\alpha  > 0$. The second part of condition (a) is proved by induction on $i$. Fix  $c \ge  0$.  Let $A_{i}(c)$,  $2 \le  i \leq n$, be a sequence defined by \Ref{eq5}. Suppose that
$m_{i}(c+\alpha ) \le  m_{i}(c) + 1$. If $a_{i+1}\neq  0$, then, by the definition of the number $\alpha$,
$$
  A_{i+1}(c+\alpha ) \le  A_{i+1}(c) + \frac{l_{i+1}}{l_{i}}+ \alpha  a_{i+1}
  \le \llc A_{i+1}(c)\rrc + 1.
$$
If $a_{i+1}= 0$, then the above equalities are also satisfied. Hence,
\begin{align*}
m_{i+1}(c+\alpha ) = &\min \left(\llc A_{i+1}(c+\alpha) \rrc, m_{i}(c+\alpha) \right)\\
\le& \min \left(\llc A_{i+1}(c)\rrc+1, m_{i}(c)+1 \right) = m_{i+1}(c)+ 1.
\end{align*}

Let $Y \in[\Phi(0), \Phi(\delta)]$. If the function  $\Phi$ is not constant on the interval $[0,\delta]$, then there is a point of discontinuity $c_k \in  [0, \delta]$ such that $\Phi(c_k) \leq Y < \Phi(c_{k}+\alpha)$. By (a), for every $2 \le  i \leq  n$,
$$
m_{i}(c_k+\alpha ) \le  m_{i}(c_k) + 1.
$$
Hence,
$$
\Phi(c_k) \leq Y < \Phi(c_k+\alpha) \leq \Phi(c_k) + n-1,
$$
which concludes (b).
\qed
\end{proof}

\begin{remark}
In order to determine a point
 $x$ satisfying assertions of Theorem 4(b)
we do not need to find the discontinuity points of $\Phi$.
The point $x$ can be found by consecutive dividing
intervals into intervals of equal length
after $\log_{2}\llc \delta/\alpha\rrc$ steps
(starting from the interval $[0,\delta]$
 and ending with the interval of length less than or equal to
 $\alpha$).
 \end{remark}

\begin{theorem}
Let  $l_{i}> 0$ be any non-increasing sequence,
 $M \in\mathbf{N}$, and
$a_{i}\ge  0$. For every $c\ge0$, let $m_i(c)$ denote the degressively proportional
sequence with respect to $l_i$, with the initial condition $(M,\{ca_i\})$.
Suppose that $\Phi(c_0)$ is not the largest value of the function $\Phi \colon{c\mapsto{\Phi_i(c)  = \sum m_i(c)}}$. Set
$$
J = J(c_0) = \{ 2 \le  i \leq n : m_{i}(c_0) < m_{i-1}(c_0) \mbox{ and } a_{i}\neq  0\},
$$
$$
\beta = \beta (c_0) = \min\left\{ \frac1{a_{i}}\left(\llc A_{i}(c_0)\rrc -A_{i}(c_0)\right): i \in J \right \},
$$
where $A_{i}(c_0)$,  $2 \le  i \leq n$, is a sequence defined by \Ref{eq5}.

Then $J \neq\emptyset$ and $c_0 + \beta$  is the first discontinuity point
belonging to $[c_0,\infty)$.
\end{theorem}

\begin{proof}
By Theorem 3(a), $J \neq\emptyset$. Let $j$ be the smallest natural number belonging to $J$ such, that
$\beta a_{j}=\llc A_{j}(c_0)\rrc-A_{j}(c_0)$. We first prove

(i)  for every  $1 \le  i \le  n$, $m_{i}(c_0+\beta ) = m_{i}(c_0)$,

(ii) if $i < j$, then  $m_{i}(c_0+\beta +\eps ) = m_{i}(c_0)$, for sufficiently small  $\eps  > 0$,

(iii) $m_{j}(c_0+\beta +\eps ) = m_{j}(c_0)+1$  for sufficiently small
$\eps  > 0$.

Proof (i).
We proceed by induction on $i$. Suppose that $m_{i}(c_0+~\beta ) = m_{i}(c_0)$.
If ${i+1} \in  J$, then, by the definition of the constant $\beta $, we obtain
$$
A_{i+1}(c_0+\beta ) = A_{i+1}(c_0) + \beta  a_{i+1}\le  \llc A_{i+1}(c_0)\rrc.
$$
Since $\llc A_{i+1}(c_0+\beta) \rrc = \llc A_{i+1}(c_0)\rrc$, we have $m_{i+1}(c_0+\beta ) = m_{i+1}(c_0)$.
In the case ${i+1} \not\in  J$, the proof is analogous to the proof of Theorem 3(a).

Proof (ii).
We proceed by induction on $i < j$.
Assume that  $i+1 < j$ and $m_{i}(c_0+\beta +\eps ) = m_{i}(c_0)$, for sufficiently small $\eps > 0$. If $i+1 \in J$, then, by the definition of the constant $\beta $  and by that of the number $j$,
 $\beta  a_{i+1}< \llc A_{i+1}(c_0)\rrc - A_{i+1}(c_0)$. Hence, for another sufficiently small  $\eps  > 0$, we obtain
$$
A_{i+1}(c_0+\beta +\eps ) = A_{i+1}(c_0) + (\beta +\eps )a_{i+1}
< \llc A_{i+1}(c_0)\rrc.
$$
Since $\llc A_{i+1}(c_0+\beta +\eps ) \rrc = \llc A_{i+1}(c_0)\rrc$, we have $m_{i+1}(c_0+\beta +\eps ) = m_{i+1}(c_0)$.
In the case $i+1 \not\in  J$, the proof is analogous to the proof of Theorem 3(a).

Proof (iii).
By (ii), $m_{j-1}(c_0+\beta +\eps ) = m_{j-1}(c_0)$, for sufficiently small
$\eps  > 0$. Hence, by the definition of the number $j$,  we have
$$
A_{j}(c_0+\beta +\eps ) = A_{j}(c_0) + (\beta +\eps )a_{j}
= \llc A_{j}(c_0)\rrc + \eps  a_{j}< \llc A_{j}(c_0)\rrc + 1.
$$
Therefore, for sufficiently small $\eps  > 0$,  we obtain
\begin{align*}
m_{j}(c_0+\beta +\eps )
= & \min \left(\llc A_{j}(c_0+\beta +\eps )\rrc, m_{j-1}(c_0)\right)\\
 =& \min \left(\llc A_{j}(c_0)\rrc+1, m_{j-1}(c_0)\right) = m_{j}(c_0) +1.
\end{align*}
The least equality follows from  $m_{j}(c_0) < m_{j-1}(c_0)$.

By  (i), $\Phi (c_0+\beta ) = \Phi (c_0)$, while by (iii),
$\Phi (c_0+\beta+\eps)> \Phi (c_0)$,  for $\eps > 0$,
and the proof is complete.
\qed
\end{proof}

\begin{remark}
If  $l_{i}$  and  $a_{i}$
are sequences of rational numbers,
then all the points of discontinuity of $\Phi  =\sum m_i$
are rational too.
\end{remark}

\begin{theorem}
Let  $l_{i}> 0$ be any non-increasing sequence,
 $M \in\mathbf{N}$, and suppose that  $a_{i}\ge  0$ is different from the zero sequence $(0,\ldots,0)$ satisfying the condition \Ref{eq4}. For every $c\ge0$, let $m_i(c)$ denote the degressively proportional
sequence with respect to $l_i$, with the initial condition $(M,\{ca_i\})$.
For  $c \ge  0$, set
$$
\begin{array}{l}
\omega = \omega(c) = \min \left\{ \dfrac1{a_{i}}\left(\llc A_{i}(c)\rrc -A_{i}(c)\right): a_{i}\neq  0 \right \},\\
J_{1} =
\left\{ 2 \le  i \leq n : A_{i}(c)) + \dfrac{l_{i}}{l_{i-1}}
<\llc A_{i}(c)\rrc \mbox{ and  $a_{i}\neq  0$} \right\}, \\
J_{2} =
\left\{ 2 \le  i \leq n : \llc A_{i}(c) \rrc
\le  A_{i}(c) + \dfrac{l_{i}}{l_{i-1}} \mbox{ and  $a_{i}\neq  0$}\right\}, \\
J_{3}=
\left\{ 2 \le  i \leq n :\dfrac1{a_{i}}\left( \llc A_{i}(c)\rrc - A_{i}(c)\right) = \omega \mbox{ and } i\in J_{2}
\right\}, \\
\gamma _{1} =
\min \left\{\dfrac1{a_{i}}\left(\llc A_{i}(c)\rrc - A_{i}(c)
- \dfrac{l_{i}}{l_{i-1}} \right): i \in  J_{1}\right\}, \\
\gamma _{2} =
\min \left\{\dfrac1{a_{i}}\left(\llc A_{i}(c)\rrc - A_{i}(c)\right) : i \in  J_{2}\backslash J_{3}
\right\}, \\
\gamma _{3} =
\min \left\{\dfrac1{a_{i}}\left(1 - \dfrac{l_{i}}{l_{i-1}}\right): i \in  J_{3}\right\}, \\
\end{array}
$$
where $A_{i}(c)$,  $2 \le  i \leq n$, is a sequence defined by \Ref{eq5}.

Then,
$$
\gamma  = \gamma (c) = \min \{\gamma _{k}: J_{k}\neq \emptyset  \} > 0 \, .
$$
The function  $\Phi \colon{c\mapsto{\Phi_i(c)  = \sum m_i(c)}}$
has in $[c,c+\gamma(c)]$ at most one point of discontinuity.
\end{theorem}

\begin{proof}
Since $a_{i} $  is different from the zero sequence, the set $J_{1}\cup  J_{2}$ is not empty. Hence, by implication \Ref{eq4},
$\gamma  > 0$. We now turn to the next part of the proof. By Theorem~5, we may assume that  $\omega(c) \leq \beta(c) < \gamma(c)$.
It suffices to show that $\Phi$ is constant at the interval $(c+\omega, c+\gamma ]$. To this purpose, we prove that one of the following conditions is satisfied for
every  $2 \le  i \leq n$:

(j) $m_{i}$ equals $m_{i}(c)$ on the interval $(c+\omega, c+\gamma ]$,

(jj) $m_{i}$ equals $m_{i}(c)+1$ on the interval $(c+\omega, c+\gamma ]$.

\noindent
We proceed by induction. Suppose that condition (j) or (jj) holds for $i$. We first prove that one of the following conditions holds for $i+1$.

(k) $\llc A_{i+1}\rrc$ equals $\llc A_{i+1}(c)\rrc$ on the interval $(c+\omega, c+\gamma ]$,

(kk) $\llc A_{i+1}\rrc$ equals $\llc A_{i+1}(c)\rrc+1$  on the interval $(c+\omega, c+\gamma ]$.

\noindent
If $i+1 \in  J_{1}$, then, for every  $x \in  (c+\omega, c+\gamma ]$,
$$
A_{i+1}(c) \le  A_{i+1}(x) \le  A_{i+1}(c) + \dfrac{l_{i+1}}{l_{i}}+ (x-c) a_{i+1}
\le \llc A_{i+1}(c)\rrc.
$$
If $i+1 \in  J_{2}$ and condition (j) (respectively (jj)) holds for $i$, then, for every $x \in  (c+\omega, c+\gamma ]$,
$$
 A_{i+1}(c) \le  A_{i+1}(x) = A_{i+1}(c) + (x-c) a_{i+1}
 \le \llc A_{i+1}(c)\rrc
$$
(or
$$
\llc A_{i+1}(c)\rrc < A_{i+1}(x)
= A_{i+1}(c) + \dfrac{l_{i+1}}{l_{i}}+ (x-c) a_{i+1}
\leq \llc A_{i+1}(c)\rrc + 1,
$$
respectively).
If $i+1 \in  J_{3}$, then, for every  $x \in  (c+\omega, c+\gamma ]$,
\begin{align*}
\llc A_{i+1}(c)\rrc &<  \llc A_{i+1}(c)\rrc + (x-c-\omega) a_{i+1} = A_{i+1}(c) + (x-c) a_{i+1} \\
 &\leq    A_{i+1}(x) \le  A_{i+1}(c) + \dfrac{l_{i+1}}{l_{i}}+ (x-c) a_{i+1} \le \llc A_{i+1}(c)\rrc + 1.
\end{align*}
If $a_{i+1}= 0$, then
$$
A_{i+1}(x)=
\left\{
\begin{array}{ll}
A_{i+1}(c)&\mbox{for  $m_{i}(x) = m_{i}(c)$},\\
 A_{i+1}(c) + \dfrac{l_{i+1}}{l_{i}}&
\mbox{for  $m_{i}(x) = m_{i}(c)+1$}.
\end{array}
\right.
$$
We proceed to show that condition (j) or (jj) holds for $i+1$.
If condition (j) holds for $i$, and condition (k) holds for $i+1$, then, for every $x \in  (c+\omega, c+\gamma ]$, we have
$$
m_{i+1}(x) =
\min \left(\llc A_{i+1}(x)\rrc, m_{i}(x)\right)
=\min \left(\llc A_{i+1}(c)\rrc, m_{i}(c)\right)
= m_{i+1}(c).
$$
If condition (jj) holds for $i$, and condition (kk) holds for $i+1$, then, for every $x \in  (c+\omega, c+\gamma ]$, we have
\begin{align*}
m_{i+1}(x) = &
\min \left(\llc A_{i+1}(x)\rrc, m_{i}(x)\right)\\
 = & \min \left(\llc A_{i+1}(c)\rrc+1, m_{i}(c)+1\right)
 = m_{i+1}(c) +1.
\end{align*}
If condition (j) holds for $i$, and condition (kk) holds for $i+1$, then, for every $x \in  (c+\omega, c+\gamma ]$, we have
$$
m_{i+1}(x) =
\min \left(\llc A_{i+1}(c)\rrc+1, m_{i}(c)\right)
=\left\{
\begin{array}{ll}
m_{i+1}(c)& \mbox{for  $\llc A_{i+1}(c)\rrc \ge  m_{i}(c)$},\\
m_{i+1}(c)+1& \mbox{for  $\llc A_{i+1}(c)\rrc < m_{i}(c)$}.
\end{array}
\right.
$$
If condition (jj) holds for $i$, and  condition (k) holds for $i+1$, then, for every $x \in  (c+\omega, c+\gamma ]$, we have
$$
 m_{i+1}(x) =
 \min \left(\llc A_{i+1}(c)\rrc, m_{i}(c)+1\right)
=\left\{
\begin{array}{ll}
m_{i+1}(c)&  \mbox{for  $\llc A_{i+1}(c)\rrc \le  m_{i}(c)$},\\
m_{i+1}(c)+1& \mbox{for  $\llc A_{i+1}(c)\rrc > m_{i}(c)$},
\end{array}
\right.
$$
which completes the proof.
\qed
\end{proof}

\begin{remark}
We conclude from Theorems~5 and 6 that if $\Phi(c)$ is not the largest value of the function $\Phi$, then
$$\Phi(c) \mbox{ and } \Phi(c+\beta(c)+\gamma(c+\beta(c)))$$
are consecutive values of $\Phi$.
\end{remark}

\section{Examples}

Recall that for a real number $x$ we denote by ${\lfloor x\rfloor}_8$ the decimal representation of $x$ up to eight digits after the decimal point.
Suppose $l_1 > l_2 > \ldots > l_{27}$ is the sequence of populations of Member States of the European Parliament. Let  $m_i(c)$, $c \geq 0$, be a degressively proportional
sequence with respect to $l_i$,  with the initial condition $(96,\{c\})$, and $\Phi(c) = \sum m_i(c)$.

Suppose we want to find $1.11 < c_2 < c_3$, such that $\Phi(1.11)$, $\Phi(c_2)$, $\Phi(c_3)$ are consecutive values of $\Phi$. This can be accomplished in the following steps :
\begin{itemize}
\item[S1.]
Set $c_1 = 1.11$. Find  ${\lfloor \beta(c_1)\rfloor}_8$ (cf. Table 3).
\item[S2.]
Set $d_1 = c_1 + {\lfloor \beta(c_1)\rfloor}_8$. Find  ${\lfloor \gamma(d_1)\rfloor}_8$ (cf. Table 4).
\item[S3.]
Set $c_2 = d_1 + {\lfloor \gamma(d_1)\rfloor}_8$. Find  ${\lfloor \beta(c_2)\rfloor}_8$ (cf. Table 5).
\item[S4.]
Set $d_2 = c_2 + {\lfloor \beta(c_2)\rfloor}_8$. Find  ${\lfloor \gamma(d_2)\rfloor}_8$ (cf. Table 6).
\item[S5.]
Set $c_3 = d_2 + {\lfloor \gamma(d_2)\rfloor}_8$. Find  ${\lfloor \beta(c_3)\rfloor}_8$ (cf. Table 7).
\item[S6.]
Set $d_3 = c_3 +{\lfloor \beta(c_3)\rfloor}_8$.
\end{itemize}
By Theorem~5, $\Phi$ is constant on the interval $[c_1, d_1]$, $[c_2, d_2]$ and $[c_3, d_3]$. Since $\Phi(d_1) < \Phi(c_2)$ and $\Phi(d_2) < \Phi(c_3)$,
Theorem 6 shows that $\Phi$ has exactly one discontinuity point in $[d_1, c_2]$, and in $[d_2, c_3]$ too.

\renewcommand{\textfraction}{0}
\newcounter{lp}
\setcounter{lp}{0}

\gdef\1#1;#2;#3;#4;#5;#6;;;;;;;;;{\stepcounter{lp} \arabic{lp}.&#1& $#2$ &$#3$ &$#4$ &$#5$ &$#6$ \\[3pt]
\hline}
\gdef\2#1{\multicolumn{1}{c}{$#1$}}
\catcode`\,=13
\def,{.}
{\tabcolsep5pt
\begin{table}
\footnotesize
\centering
\begin{tabular}{rl*6{r}}
\hline
 && \2{l_i}&\2{A_i}&\2{\llc A_i\rrc}&\2{M_i}&\2{\rule{0pt}{10pt}1-\dfrac{l_i}{l_{i-1}}}\\[3pt]
 \hline
\1\DE;82,438;;;96;;;;;;;;;;
\1\FR;62,999;73,36306072;74;74;0,235801451;;;;;;;;;
\1\GB;60,393;70,93893554;71;71;0,041365736;;;;;;;;;
\1\IT;58,752;69,07078635;70;70;0,027172023;;;;;;;;;
\1\ES;43,758;52,13541667;53;53;0,255208333;;;;;;;;;
\1\PL;38,157;46,21602907;47;47;0,127999452;;;;;;;;;
\1\RO;21,610;26,61818277;27;27;0,433655686;;;;;;;;;
\1\NL;16,334;20,40805183;21;21;0,244146229;;;;;;;;;
\1\GR;11,125;14,30298763;15;15;0,318905351;;;;;;;;;
\1\PT;10,570;14,25168539;15;15;0,049887640;;;;;;;;;
\1\BE;10,511;14,91627247;15;15;0,005581835;;;;;;;;;
\1\CZ;10,251;14,62896014;15;15;0,024735991;;;;;;;;;
\1\HU;10,077;14,74539069;15;15;0,016973954;;;;;;;;;
\1\SE;9,048;13,46829414;14;14;0,102113724;;;;;;;;;
\1\AT;8,266;12,79000884;13;13;0,086427940;;;;;;;;;
\1\BG;7,719;12,13972901;13;13;0,066174692;;;;;;;;;
\1\DK;5,427;9,139914497;10;10;0,296929654;;;;;;;;;
\1\SK;5,389;9,929979731;10;10;0,007002027;;;;;;;;;
\1\FI;5,256;9,753200965;10;10;0,024679904;;;;;;;;;
\1\IE;4,209;8,007990868;9;9;0,199200913;;;;;;;;;
\1\LT;3,403;7,276550249;8;8;0,191494417;;;;;;;;;
\1\LV;2,295;5,395239495;6;6;0,325595063;;;;;;;;;
\1\SI;2,003;5,236601307;6;6;0,127233115;;;;;;;;;
\1\EE;1,345;4,028956565;5;5;0,328507239;;;;;;;;;
\1\CY;0,766;2,847583643;3;3;0,430483271;;;;;;;;;
\1\LU;0,469;1,836814621;2;2;0,387728460;;;;;;;;;
\1\MT;0,405;1,727078891;2;2;0,136460554;;;;;;;;;
&$\Phi(0)$\rule{0pt}{10pt}&&&&645&\\
\hline
\end{tabular}
\catcode`\,=12
\caption{In column 5 of Table 2 the values of minorant $M_i$ are put, cf.~Theorem~1(c). Note that ${\lfloor 1 - l_{i}/l_{i-1} \rfloor}_8 > 0$, for $2 \leq  i \leq 27$.
}
\end{table}

\setcounter{lp}{0}
\def\1#1;#2;#3;#4;#5;#6;;;;;;;{\stepcounter{lp} \arabic{lp}.&#1& $#2$ &$#3$ &$#4$ &$#5$ &$#6$ \\[3pt]
\hline}
\catcode`\,=13
\def,{.}
\def\3#1{ && \2{l_i}&\2{A_i}&\2{\llc A_i\rrc}&\2{m_i(c_{#1})}&\2{\rule{0pt}{10pt}\llc A_i\rrc-A_i}\\
\hline}
\def\4#1{ && \2{l_i}&\2{A_i}&\2{\llc A_i\rrc}&\2{m_i(d_{#1})}&\2{\rule{0pt}{12pt}\llc A_i\rrc-A_i}&
\2{\llc A_i\rrc{-}A_i {-} \dfrac{l_i}{l_{i-1}}}\\
\hline}
\begin{table}
\centering
\begin{tabular}{rl*5{r}}
\hline
\31
\1\DE;82,438;;;96;;;;;;;;
\1\FR;62,999;74,47306072;75;75;0,526939276;;;;;;;
\1\GB;60,393;73,0075698;74;74;0,992430197;;;;;;;
\1\IT;58,752;73,09927028;74;74;0,900729720;;;;;;;
\1\ES;43,758;56,22458333;57;57;0,775416667;;;;;;;
\1\PL;38,157;50,81403126;51;51;0,185968737;;;;;;;
\1\RO;21,610;29,99356003;30;30;\mathit{0,006439972};;;;;;;
\1\NL;16,334;23,78561314;24;24;0,214386858;;;;;;;
\1\GR;11,125;17,45627158;18;18;0,543728419;;;;;;;
\1\PT;10,570;18,21202247;19;18;0,787977528;;;;;;;
\1\BE;10,511;19,00952696;20;18;0,990473037;;;;;;;
\1\CZ;10,251;18,66475216;19;18;0,335247836;;;;;;;
\1\HU;10,077;18,80446883;19;18;0,195531168;;;;;;;
\1\SE;9,048;17,27195296;18;18;0,728047038;;;;;;;
\1\AT;8,266;17,55429708;18;18;0,445702918;;;;;;;
\1\BG;7,719;17,91885555;18;18;0,081144447;;;;;;;
\1\DK;5,427;13,76526623;14;14;0,234733774;;;;;;;
\1\SK;5,389;15,01197162;16;14;0,988028377;;;;;;;
\1\FI;5,256;14,76448135;15;14;0,235518649;;;;;;;
\1\IE;4,209;12,32118721;13;13;0,678812785;;;;;;;
\1\LT;3,403;11,62057258;12;12;0,379427417;;;;;;;
\1\LV;2,295;9,202859242;10;10;0,797140758;;;;;;;
\1\SI;2,003;9,837668845;10;10;0,162331155;;;;;;;
\1\EE;1,345;7,824927609;8;8;0,175072391;;;;;;;
\1\CY;0,766;5,666133829;6;6;0,333866171;;;;;;;
\1\LU;0,469;4,783629243;5;5;0,216370757;;;;;;;
\1\MT;0,405;5,427697228;6;5;0,572302772;;;;;;;
& $\Phi(c_1)$\rule{0pt}{10pt}&&&&736&\\
\hline
\end{tabular}
\catcode`\,=12
\caption{
$c_1 = 1.11$ and ${\lfloor \beta(c_1) \rfloor}_8 = 0.00643997$.
In column~6 of Table 3 the values of the sequence $m_i(c_1)$ are put.
}
\end{table}

\setcounter{lp}{0}
\def\1#1;#2;#3;#4;#5;#6;#7;;;;;;{\stepcounter{lp} \arabic{lp}.\rule{0pt}{10pt}&#1& $#2$ &$#3$ &$#4$ &$#5$ &$#6$ &$#7$\\[3pt]
\hline}
\catcode`\,=13
\def,{.}
\begin{table}
\scriptsize
\centering
\begin{tabular}{rl*6{r}}
\hline
\41
\1\DE;82,438;;96;96;;;;;;;;
\1\FR;62,999;74,47950069;75;75;0,520499306;-0,243699244;;;;;;
\1\GB;60,393;73,01400977;74;74;0,985990227;0,027355963;;;;;;
\1\IT;58,752;73,10571025;74;74;0,894289750;-0,078538226;;;;;;
\1\ES;43,758;56,23102330;57;57;0,768976697;\mathit{0,024185030};;;;;;
\1\PL;38,157;50,82047123;51;51;0,179528767;-0,692471781;;;;;;
\1\RO;21,610;30,00000000;30;30;1,69589E{-09};-0,566344313;;;;;;
\1\NL;16,334;23,79205311;24;24;0,207946888;-0,547906883;;;;;;
\1\GR;11,125;17,46271155;18;18;0,537288449;-0,143806200;;;;;;
\1\PT;10,570;18,21846244;19;18;0,781537558;-0,168574801;;;;;;
\1\BE;10,511;19,01596693;20;18;0,984033067;-0,010385098;;;;;;
\1\CZ;10,251;18,67119213;19;18;0,328807866;-0,646456144;;;;;;
\1\HU;10,077;18,81090880;19;18;0,189091198;-0,793934849;;;;;;
\1\SE;9,048;17,27839293;18;18;0,721607068;-0,176279208;;;;;;
\1\AT;8,266;17,56073705;18;18;0,439262948;-0,474309112;;;;;;
\1\BG;7,719;17,92529552;18;18;\mathit{0,074704477};-0,859120831;;;;;;
\1\DK;5,427;13,77170620;14;14;0,228293804;-0,474776542;;;;;;
\1\SK;5,389;15,01841159;16;14;0,981588407;-0,011409566;;;;;;
\1\FI;5,256;14,77092132;15;14;0,229078679;-0,746241417;;;;;;
\1\IE;4,209;12,32762718;13;13;0,672372815;-0,128426271;;;;;;
\1\LT;3,403;11,62701255;12;12;0,372987447;-0,435518136;;;;;;
\1\LV;2,295;9,209299212;10;10;0,790700788;0,116295851;;;;;;
\1\SI;2,003;9,844108815;10;10;0,155891185;-0,716875700;;;;;;
\1\EE;1,345;7,831367579;8;8;0,168632421;-0,502860339;;;;;;
\1\CY;0,766;5,672573799;6;6;0,327426201;-0,242090528;;;;;;
\1\LU;0,469;4,790069213;5;5;0,209930787;-0,402340753;;;;;;
\1\MT;0,405;5,434137198;6;5;0,565862802;-0,297676644;;;;;;
&$\Phi(d_1)$\rule{0pt}{8pt}&&&&736&&\\
\hline
\end{tabular}
\catcode`\,=12
\caption{$
d_1 = c_1+{\lfloor \beta(c_1) \rfloor}_8 = 1.11643997$.
Since $J_3 = \{7\}$ and  ${\lfloor 1 - l_{7}/l_{6} \rfloor}_8 =0.43365568$ (see Table 2), we have
${\lfloor \gamma(d_1)\rfloor}_8 = 0.02418503$.}
\end{table}

\setcounter{lp}{0}
\def\1#1;#2;#3;#4;#5;#6;;;;;;;{\stepcounter{lp} \arabic{lp}.&#1& $#2$ &$#3$ &$#4$ &$#5$ &$#6$ \\[3pt]
\hline}
\catcode`\,=13
\def,{.}
\begin{table}
\centering
\begin{tabular}{rl*5{r}}
\hline
\32
\1\DE;82,438;;;96;;;;;;;;
\1\FR;62,999;74,50368572;75;75;0,496314276;;;;;;;
\1\GB;60,393;73,03819480;74;74;0,961805197;;;;;;;
\1\IT;58,752;73,12989528;74;74;0,870104720;;;;;;;
\1\ES;43,758;56,25520833;57;57;0,744791667;;;;;;;
\1\PL;38,157;50,84465626;51;51;0,155343737;;;;;;;
\1\RO;21,610;30,02418503;31;31;0,975814972;;;;;;;
\1\NL;16,334;24,57209191;25;25;0,427908087;;;;;;;
\1\GR;11,125;18,16799123;19;19;0,832008770;;;;;;;
\1\PT;10,570;19,19275983;20;19;0,807240169;;;;;;;
\1\BE;10,511;20,03457013;21;19;0,965429872;;;;;;;
\1\CZ;10,251;19,67064117;20;19;0,329358826;;;;;;;
\1\HU;10,077;19,81811988;20;19;0,181880121;;;;;;;
\1\SE;9,048;18,20046424;19;19;0,799535762;;;;;;;
\1\AT;8,266;18,49849414;19;19;0,501505858;;;;;;;
\1\BG;7,719;18,88330586;19;19;0,116694139;;;;;;;
\1\DK;5,427;14,49896157;15;15;0,501038428;;;;;;;
\1\SK;5,389;16,03559460;17;15;0,964405404;;;;;;;
\1\FI;5,256;15,77042645;16;15;0,229573553;;;;;;;
\1\IE;4,209;13,15261130;14;14;0,847388699;;;;;;;
\1\LT;3,403;12,45970317;13;13;0,540296834;;;;;;;
\1\LV;2,295;9,907889179;10;10;\mathit{0,092110821};;;;;;;
\1\SI;2,003;9,868293845;10;10;0,131706155;;;;;;;
\1\EE;1,345;7,855552609;8;8;0,144447391;;;;;;;
\1\CY;0,766;5,696758829;6;6;0,303241171;;;;;;;
\1\LU;0,469;4,814254243;5;5;0,185745757;;;;;;;
\1\MT;0,405;5,458322228;6;5;0,541677772;;;;;;;
&$\Phi(c_2)$\rule{0pt}{10pt}&&&&751&\\
\hline
\end{tabular}
\catcode`\,=12
\caption{$
c_2 = d_1 + {\lfloor \gamma(d_1)\rfloor}_8 = 1.140625 \hbox{ and } {\lfloor \beta(c_2)\rfloor}_8 = 0.09211082.
$
In column 6 of Table~5 the values of the sequence $m_i(c_2)$ are put.
}
\end{table}

\setcounter{lp}{0}
\def\1#1;#2;#3;#4;#5;#6;#7;;;;;{\stepcounter{lp} \arabic{lp}.\rule{0pt}{10pt}&#1& $#2$ &$#3$ &$#4$ &$#5$ &$#6$ &$#7$\\[3pt]
\hline}
\catcode`\,=13
\def,{.}
\begin{table}
\scriptsize
\centering
\begin{tabular}{rl*6{r}}
\hline
\42
\1\DE;82,438;;;96;;;;;;;
\1\FR;62,999;74,59579654;75;75;0,404203456;-0,359995094;;;;;
\1\GB;60,393;73,13030562;74;74;0,869694377;-0,088939887;;;;;
\1\IT;58,752;73,22200610;74;74;0,777993900;-0,194834076;;;;;
\1\ES;43,758;56,34731915;57;57;0,652680847;-0,092110820;;;;;
\1\PL;38,157;50,93676708;51;51;0,063232917;-0,808767631;;;;;
\1\RO;21,610;30,11629585;31;31;0,883704152;0,317359837;;;;;
\1\NL;16,334;24,66420273;25;25;0,335797267;-0,420056505;;;;;
\1\GR;11,125;18,26010205;19;19;0,739897950;\mathit{0,058803301};;;;;
\1\PT;10,570;19,28487065;20;19;0,715129349;-0,234983011;;;;;
\1\BE;10,511;20,12668095;21;19;0,873319052;-0,121099112;;;;;
\1\CZ;10,251;19,76275199;20;19;0,237248006;-0,738016003;;;;;
\1\HU;10,077;19,91023070;20;19;0,089769301;-0,893256745;;;;;
\1\SE;9,048;18,29257506;19;19;0,707424942;-0,190461334;;;;;
\1\AT;8,266;18,59060496;19;19;0,409395038;-0,504177022;;;;;
\1\BG;7,719;18,97541668;19;19;\mathit{0,024583319};-0,909241990;;;;;
\1\DK;5,427;14,59107239;15;15;0,408927608;-0,294142738;;;;;
\1\SK;5,389;16,12770542;17;15;0,872294584;-0,120703390;;;;;
\1\FI;5,256;15,86253727;16;15;0,137462733;-0,837857364;;;;;
\1\IE;4,209;13,24472212;14;14;0,755277879;-0,045521208;;;;;
\1\LT;3,403;12,55181399;13;13;0,448186014;-0,360319569;;;;;
\1\LV;2,295;9,999999999;10;10;1,33412E{-09};-0,674404935;;;;;
\1\SI;2,003;9,960404665;10;10;0,039595335;-0,833171550;;;;;
\1\EE;1,345;7,947663429;8;8;0,052336571;-0,619156189;;;;;
\1\CY;0,766;5,788869649;6;6;0,211130351;-0,358386378;;;;;
\1\LU;0,469;4,906365063;5;5;0,093634937;-0,518636603;;;;;
\1\MT;0,405;5,550433048;6;5;0,449566952;-0,413972494;;;;;
&$\Phi(d_2)$\rule{0pt}{8pt}&&&&751&&\\
\hline
\end{tabular}
\catcode`\,=12
\caption{$
d_2 = c_2+{\lfloor \beta(c_2)\rfloor}_8 = 1.23273582.
$
Since $J_3 = \{22\}$  and  ${\lfloor 1 - l_{22}/l_{21} \rfloor}_8 =0.32559506$ (see Table 2), we have
${\lfloor \gamma(d_2) \rfloor}_8 = 0.02458331$.}
\end{table}

\setcounter{lp}{0}
\def\1#1;#2;#3;#4;#5;#6;{\stepcounter{lp} \arabic{lp}.&#1& $#2$ &$#3$ &$#4$ &$#5$ &$#6$ \\[3pt]
\hline}
\catcode`\,=13
\def,{.}
\begin{table}
\centering
\begin{tabular}{rl*5{r}}
\hline
\33
\1\DE;82,438;;;96;;
\1\FR;62,999;74,62037985;75;75;0,379620146;
\1\GB;60,393;73,15488893;74;74;0,845111067;
\1\IT;58,752;73,24658941;74;74;0,753410590;
\1\ES;43,758;56,37190246;57;57;0,628097537;
\1\PL;38,157;50,96135039;51;51;\mathit{0,038649607};
\1\RO;21,610;30,14087916;31;31;0,859120842;
\1\NL;16,334;24,68878604;25;25;0,311213957;
\1\GR;11,125;18,28468536;19;19;0,715314640;
\1\PT;10,570;19,30945396;20;19;0,690546039;
\1\BE;10,511;20,15126426;21;19;0,848735742;
\1\CZ;10,251;19,78733530;20;19;0,212664696;
\1\HU;10,077;19,93481401;20;19;0,065185991;
\1\SE;9,048;18,31715837;19;19;0,682841632;
\1\AT;8,266;18,61518827;19;19;0,384811728;
\1\BG;7,719;18,99999999;19;19;8,64021E{-09};
\1\DK;5,427;14,61565570;15;15;0,384344298;
\1\SK;5,389;16,15228873;17;15;0,847711274;
\1\FI;5,256;15,88712058;16;15;0,112879423;
\1\IE;4,209;13,26930543;14;14;0,730694569;
\1\LT;3,403;12,57639730;13;13;0,423602704;
\1\LV;2,295;10,02458331;11;11;0,975416691;
\1\SI;2,003;10,85775486;11;11;0,142245140;
\1\EE;1,345;8,643739499;9;9;0,356260501;
\1\CY;0,766;6,382969688;7;7;0,617030312;
\1\LU;0,469;5,543219913;6;6;0,456780087;
\1\MT;0,405;6,438555804;7;6;0,561444196;
&$\Phi(c_3)$\rule{0pt}{10pt}&&&&757&\\
\hline
\end{tabular}
\catcode`\,=12
\caption{$
c_3 = d_2 + {\lfloor \gamma(d_2) \rfloor}_8 = 1.25731913.
$
In column~6 of Table 7 the values of the sequence $m_i(c_3)$ are put. Since $m_{16}(c_3) = m_{15}(c_3)$, we have
$
{\lfloor \beta(c_3) \rfloor}_8 = 0.0386496.
$
}
\end{table}

\clearpage

\end{document}